\documentstyle[12pt,epsbox]{article}
\newcommand{\square}{\kern1pt\vbox{\hrule height 1.2pt\hbox{\vrule width
1.2pt\hskip 3pt
   \vbox{\vskip 6pt}\hskip 3pt\vrule width 0.6pt}\hrule height 0.6pt}\kern1pt}
\newcommand{\lsim}{\mbox{$\;\lower-.2ex\hbox{$\textstyle<$}\;\!\!\!\!\!\!
\lower.7ex\hbox{$\textstyle \sim$}\;$}}
\newcommand{\gsim}{\mbox{$\;\lower-.2ex\hbox{$\textstyle>$}\;\!\!\!\!\!\!
\lower.7ex\hbox{$\textstyle \sim$}\;$}}
\newcommand{\rh}{r_{\scriptscriptstyle H}}
\newcommand{\OmH}{\Omega_{\scriptstyle H}}

\newcommand{\hspc}{\hspace*{\parindent}}

\begin{document}

\renewcommand{\Large}{\normalsize}
\renewcommand{\huge}{\normalsize}

\begin{titlepage}
\baselineskip .15in
\begin{flushright}
WU-AP/43/94 \\
hep-th/9408084
\end{flushright}

{}~\\

\vskip 1.5cm
\begin{center}
{\bf

\vskip 1.5cm
{\large Superradiance around Rotating Dilatonic Black Holes}

}\vskip .8in

{\sc  Jun-ichirou Koga }  and
{\sc  Kei-ichi Maeda}$^{(a)}$\\[1em]
 {\em Department of Physics, Waseda University,
Shinjuku-ku, Tokyo 169, Japan}

\end{center}
\vfill
%%%%%%%%%%%%%%%%%%%%%%%%%%%%%%%%%%%%%%%%%%%%%%%%%%%%%%%
 \begin{abstract}
We consider a superradiance effect around  rotating
dilatonic black holes.  We analyze two cases: one is
an exact solution with the coupling constant $\alpha=\sqrt{3}$,
 which effective action
 is reduced from the 5-dimensional Kaluza-Klein theory,
and the other is a slowly rotating
 dilatonic black holes with arbitrary coupling
constant.
We find that there exists a critical value ($\alpha \sim 1$), which is
 predicted from a superstring model, and
the superradiant emission rate with coupling larger than
the critical value
becomes much higher than the Kerr-Newman case
($\alpha=0$) in the maximally charged
limit.  Consequently,
 4-dimensional primordial black holes in
higher dimensional unified theories are either
rotating but  almost neutral or charged but  effectively
non-rotating.
 \end{abstract}
%%%%%%%%%%%%%%%%%%%%%%%%%%%%%%%%%%%%%%%%%%%%%%%%%%%%%%%%%%

\vfill
\begin{center}
July, 1994
\end{center}
\vfill
(a)~~electronic mail :  maeda@cfi.waseda.ac.jp\\
\end{titlepage}

\baselineskip .3in

\vspace{.5 cm}
\normalsize
\baselineskip = 24pt
%%%%%%%%%%%%%%%%%%%%%%%%%%%%%%%%%%%%%%%%%%%%%%%%%%%%%%%%%%%%%%
\section{Introduction}
\hspc
 In unified theories of fundamental interactions, a dilaton
field usually appears and couples to other known matter field.  It may play
some important role
in the history of the early universe or in astrophysics.
It may also provide
us a window to see some evidence of higher-dimensional unification.
 The coupling constant of the dilaton field
depends on the dimension of an internal space and a unification theory.
Therefore it is important to
see an effect of the coupling on observable phenomena
and to investigate a dependence of the coupling constant.

In the Einstein-Maxwell-dilaton theory,  which  action is described as
\begin{equation}
S = \frac{1}{16 \pi} \int d^{4} x \sqrt{- g} \left[ R - 2 \left( \nabla \phi
\right)^{2} - e^{- 2 \alpha \phi} F^{2} \right] ,
 \label{eqn:dilaction}
\end{equation}
electrically (or magnetically) charged non-rotating  black hole\cite{CDBsol}
shows very interesting thermodynamical properties depending on a coupling
constant $ \alpha $ of a dilaton field $ \phi $. (We adopt the units of
$c=\hbar=G=1$.) Remind
that the case of $ \alpha = 0 $  corresponds to the Einstein-Maxwell theory, in
which the Reissner-Nordstr\"om solution is a unique spherically symmetric black
hole.
If $ \alpha < 1 $, the thermodynamical property is similar to the
Reissner-Nordstr\"om black hole, namely, its Hawking temperature vanishes in
the limit of extremely charged  hole.
When we consider a superstring theory ($ \alpha = 1 $), its temperature  in the
extreme limit becomes  non-zero and finite, while
 it diverges in the case of $ \alpha > 1 $.
The last case includes the 5-dimensional Kaluza-Klein theory, where
$ \alpha = \sqrt{3} $.
This property of dilatonic black holes indicates that
the Hawking radiation may be drastically affected by the  coupling constant.

Shiraishi\cite{SuperradQ} also discussed a
discharge process by the superradiant modes for  non-rotating dilatonic black
holes.
Although he could not present a definite conclusion
because of his analytic WKB approach, his analysis suggests
that there is a critical value of the coupling constant.
Below the critical value,  the superradiant effect is similar to that of the
Reissner-Nordstr\"{o}m black hole, while
it changes qualitatively beyond that value, i.e., as the coupling constant
increases, the discharge proceeds very rapidly.\\
\hspc
 In this letter,  we study  the rotating black holes and
investigate the effect of dilaton field on the superradiance.
We also
look for whether there is a critical value beyond which the
property changes qualitatively.

One difficulty in our study is that  we do not know the exact solutions of
rotating black holes except for two cases: $ \alpha = 0 $ (Kerr-Newman) and $
\alpha = \sqrt{3} $.
The latter is the 4-dimensional rotating black hole solution in the
5-dimensional Kaluza-Klein theory derived by Frolov et al \cite{rotKKBHsol}.
We call it the Kaluza-Klein black hole.
We have, however,  the approximate solutions of  slowly rotating black holes
with arbitrary coupling constant, which were found by  Horne and Horowitz
\cite{Horowitz}, and  by Shiraishi\cite{Shiraishi}.   \\
\hspc
Here, we use these two exact solutions and the approximate solutions to analyze
superradiance around   rotating dilatonic black holes. We assume a charge of a
black hole is conserved (it is true for a central charge), and then consider
only a massless neutral scalar particle.

\section{Superradiance around Kaluza-Klein Black Hole}
\hspc
 Before showing the  calculations of  superradiance and spontaneous emission
rate, we have to mention the rotating Kaluza-Klein black hole
solution\cite{rotKKBHsol}. It is given by
\begin{eqnarray}
 ds^{2} & = &- \frac{\Delta -a^2 \sin^2 \theta}{B
\Sigma} dt^{2} - 2 a \sin^{2} \theta \frac{1}{
	      \sqrt{1-v^{2}}} \frac{Z}{B} dt d \varphi  \nonumber \\
	&   & + \left[ B \left( r^{2} + a^{2} \right) + a^{2} \sin^{2} \theta
	  \;    \frac{Z}{B} \right] \sin^{2} \theta \; d \varphi^{2} +
          \frac{B \Sigma}{\Delta} dr^{2} + B \Sigma \, d \theta^{2}
           \nonumber  \\
 A_{t}  & = &\frac{v}{2 \left( 1-v^{2} \right) } \frac{Z}{B^{2}}, \; \; \;
 A_{ \varphi } = \; - a \sin^{2} \theta \frac{v}{2 \sqrt{1-v^{2}}}
\frac{Z}{B^{2}}, \; \; \;
 \phi = \; - \frac{\sqrt{3}}{2} \ln B   \label{eqn:rotKKBHsol}
\end{eqnarray}
where, $ \Delta $, $ \Sigma $, $ Z $ and $ B $ are functions of $ r $ and $
\theta $, and given by,
\begin{equation}
 \Delta = r^{2} - 2 \mu r + a^{2}, \; \; \;
 \Sigma = r^{2} + a^{2} \cos^{2} \theta, \; \; \;
 Z = \frac{2 \mu r}{\Sigma}, \; \; \;
 B = \left( 1 + \frac{v^{2} Z}{1-v^{2}} \right)^{\frac{1}{2}}   .
\label{eqn:funcdefKK}
\end{equation}
The mass $ M $ ,  the electric charge $ Q $ and
the angular momentum $ J $ of black hole are
obtained by
the parameters $ v $ ($ |v| < 1 $) , $ \mu $ and $ a $ as,
\begin{equation}
 M=\mu\left[1 +  {v^2  \over 2(1-v^2)} \right], \; \; \;
 Q= {\mu v \over 1-v^2},
 \; \; \;
 J = {\mu a \over \sqrt{1-v^2}}
  . \label{eqn:parameterKK}
\end{equation}
Here we  assume $ J $ is positive without loss of  generality.
The horizon radius $ \rh $ is given by,
\begin{equation}
 \rh = \mu + \sqrt{ \mu^{2} - a^{2}}  ,   ~~~(\mu^2 \geq a^2)
\label{eqn:horizon}
\end{equation}
This black hole has three independent global hairs, that is, the mass $ M $,
the charge $ Q $ and the angular momentum $ J $.
When $ Q = 0 $, it reduces to the Kerr black hole, while the solution with $ J
= 0 $ corresponds to
the non-rotating dilatonic black hole \cite{CDBsol} with
the coupling constant $ \alpha = \sqrt{3} $.
The condition of $\mu^2 \geq a^2$ for the horizon to exist is rewritten as
\begin{equation}
  \left({J \over M^2}\right)^2 \leq \frac{1}{4}
\left[ 2- 10 \left({Q \over M} \right)^2 - \left({Q \over M} \right)^4
+ 2 \left(1+ 2 \left({Q \over M} \right)^2 \right)^{3/2}
\right].
\end{equation}
(see Fig. 1).
It should be noticed here that the solution
 with $ |Q| = 2 M $ (and $ J =0$)
is not a black hole solution because a naked singularity appears,
as is pointed out in \cite{CDBsol,sphKKBHsol}.  \\

 The angular velocity $ \OmH $ of this black hole is
\begin{equation}
 \OmH =  \frac{a\sqrt{1 - v^{2}}}{\rh^{2} + a^{2}}   ,
    \label{eqn:angvelocity}
\end{equation}
which shows an interesting feature near $ |Q| = 2 M $, similar to that of the
black-hole temperature\cite{CDBsol}.
When $ J = 0 $, obviously $\OmH$  vanishes. While, if $ J \neq 0 $, $ \OmH
\rightarrow \infty $ in the limit of $ |Q| \rightarrow 2M $ ($ J \rightarrow 0
$).  It is discontinuous at $ |Q| = 2 M $.
Although $ |Q| = 2 M $ is not the black hole solution, the solution with
$ |Q| \lsim 2M $ represents a rapidly spinning black hole.  \\

\hspc
Now we discuss a superradiance.
We consider a massless neutral scalar field $ \Phi $,
described by the Klein-Gordon equation
\begin{equation}
 {\large \square} \Phi = 0 \; \; .   \label{eqn:KGeqn}
\end{equation}
In the background fields of the rotating Kaluza-Klein black hole, we can
separate variables.  The spheroidal equation is written in the same form  as
the Kerr-Newman case, setting
\begin{equation}
 \Phi = \frac{\chi (r)}{R(r)} S \left(
 \theta \right) e^{ i m \varphi} e^{-i \omega t}       \label{eqn:SepPsi}
\end{equation}
where
\begin{equation}
 R^{2}(r) \equiv  B \Sigma |_{\theta =0} =
(r^{2} + a^{2})  \left( 1 + \frac{v^{2}}{1-v^{2}}
\frac{2 \mu r}{r^{2} + a^{2}}
   \right)^{\frac{1}{2}}  .  \label{eqn:B0def}
\end{equation}
\hspc
We perform a coordinate transformation from $ r $ to $ r^* $
\begin{equation}
 d r^* = \frac{R^{2}(r)}{\Delta (r)} \; d r   ,
 \label{eqn:transcoord}
\end{equation}
to get the radial equation in a usual form as
\begin{equation}
 \left[ \frac{d^{2}}{d r^{* 2}}
+ \left(\omega^{2} -  m \Omega (r) \right)^2
- V^2(r)  \right]
   \chi  \left( r^* \right) = 0  ,
  \label{eqn:radialeqn}
\end{equation}
where
\begin{eqnarray}
\Omega(r) & \equiv & {a \over \sqrt{1-v^2}}{2 \mu r
\over R^4(r)}\\
V^2(r) & \equiv &  {\Delta (r) \over R^2 (r) }
\left\{ \frac{\lambda }{R^{2}(r)}
  + \frac{1}{R(r)} \frac{d}{dr}
\left[ \frac{\Delta (r)}{R^2(r)}
\frac{d  R(r)}{dr} \right]
- {m^2 a^2 \over R^6 (r) }
\left[ r^2 +a^2 +{2 \mu r \over 1-v^2} \right]
\right\}
\end{eqnarray}

The eigenvalue
$\lambda $ of the spheroidal equation for
the angular function $S(\theta)$ is
obtained perturbatively \cite{eigenvalue} as
\begin{equation}
 \lambda = l(l+1) + \lambda_{2}(l,m) a^{2} \omega^{2} +
 \lambda_{4}(l,m) a^{4}
    \omega^{4} + { \cal O} \left( a^{6} \omega^{6} \right) \; .
   \label{eqn:eigenvalue}
\end{equation}
where $\lambda_{2}$ and $\lambda_{4}$ are some known functions of $l$ and $m$,
and $\lambda_{2}(1,1)=4/5$ and $\lambda_{2}(1,1)=-4/875$.
To see that this expansion is sufficiently powerful to
calculate $\lambda$, for example, we can point out the fact that
$ a \omega \leq 0.5 $ for the present model with $ m = 1 $, which equality
happens in the extremely rotating case (in the  Kerr limit).   \\  \hspc
 From Eq. (12), we find that a superradiance occurs for $0< \omega
< \Omega_H$.   When we quantize a field, we naturally expect spontaneous
emission for the superradiant modes, by which a rotating black hole loses the
energy and angular momentum\cite{Unruh}.    The rates of the spontaneous
emission for a scalar field is calculated by the formulae
\begin{equation}
 \frac{d M}{d t} = - \frac{1}{2 \pi} \sum_{l, m}
\int^{m \OmH}_{0}
          \omega \left( | A |^{2} -1 \right) d \omega, \; \;        \;
 \frac{d J}{d t} = - \frac{1}{2 \pi} \sum_{l, m}  \int^{m \OmH}_{0}
          m  \left( | A |^{2} -1 \right) d \omega   .
    \label{eqn:flux}
\end{equation}
where $ |A|^{2} $ is a reflection coefficient of scalar wave described by
Eq.(\ref{eqn:radialeqn}) .
\\  \hspc
In what follow, we assume the charge of the black hole is positive ($ v \geq 0
$)
without loss of generality.
We consider a process in which a charge of the black hole is conserved.
We will show our numerical results only for $ l=m=1 $ mode which is the most
dominant. \\  \hspc
 Remind that a highly rotating case is similar to the Kerr black hole.
Hence  we shall restrict our analysis  mainly into  a
highly charged case ( $ Q \lsim 2 M $).
The result of the Kaluza-Klein black hole ($\alpha=\sqrt{3}$) is shown in Fig.2
for $ J = 0.01 M^{2} $ with  the Kerr-Newman case ($\alpha=0$) to see the
effect of the dilaton coupling.
In the Kerr-Newman case, both energy and angular momentum emission rates first
increase when the charge of the black hole becomes large, and then they drop to
small values near the extreme hole.
On the other hand, both rates in the Kaluza-Klein black hole
always increase and seem to diverge to infinity in the limit of the maximal
black hole.   Therefore, we may conclude that
the superradiant emission is affected very much by the dilaton coupling, and
the emission rates in the Kaluza-Klein black hole are much higher than the
Kerr-Newman case, in particular in the extreme
limit.

\section{Superradiance around Slowly Rotating Black Holes}
\hspc
Next, we analyze a dependence of the emission rates on the coupling constant
$\alpha$. To see this, we consider the slowly rotating  solution. This solution
is described as \cite{Horowitz},\cite{Shiraishi}
\begin{eqnarray}
ds^{2} & = & -{\Delta(\tilde{r}) \over {R^2(\tilde{r})}}
dt^{2} + \frac{{R^2(\tilde{r})}}{\Delta(\tilde{r})} d\tilde{r}^{2} + {
{R}^{2}(\tilde{r})} (d \theta^{2} + \sin^{2} \theta d \varphi^{2} ) - 2 a
f(\tilde{r}) \sin^{2} \theta  d t d \varphi \nonumber \\
A_{t} & = & \frac{Q}{\tilde{r}}, \; \; \; A_{\varphi} = - a \sin^{2} \theta
\frac{Q}{\tilde{r}}, \; \; \; \phi = \frac{\alpha}{1+\alpha^{2}} \ln \left(1 -
\frac{\tilde{r}_{-}}{\tilde{r}} \right) .   \label{eqn:slowsol}
\end{eqnarray}
where we  introduced new radial coordinate $\tilde{r}~$
by a transformation, $r=\tilde{r}-\tilde{r}_-$, and
\begin{eqnarray}
 \Delta (\tilde{r}) & = & \left( \tilde{r} - \tilde{r}_{+} \right) \left(
\tilde{r} -
  \tilde{r}_{-} \right),
        \;  \; \;
   R (\tilde{r}) = \tilde{r} \left( 1 - \frac{\tilde{r}_{-}}{\tilde{r}}
\right)^{\alpha^{2}/(1  +\alpha^{2})}   \nonumber  \\
 f (\tilde{r}) & = & \frac{\left( 1+\alpha^{2} \right)^{2}}{\left( 1-\alpha^{2}
  \right) \left( 1-3\alpha^{2} \right) } \left({\tilde{r} \over
\tilde{r}_{-}}\right)^2
\left( 1 -   \frac{\tilde{r}_{-}}{\tilde{r}}
\right)^{2\alpha^{2}/(1  +\alpha^{2})}
      \nonumber  \\
    & - &   \left( 1 + \frac{\left( 1+\alpha^{2} \right)^{2}}{\left(
  1-\alpha^{2} \right) \left( 1-3\alpha^{2} \right) }
 \left({\tilde{r} \over \tilde{r}_{-}}\right)^2
  + \frac{1+\alpha^{2}}{\left( 1-\alpha^{2} \right) } \left({\tilde{r} \over
\tilde{r}_{-}}\right) -   \frac{\tilde{r}_{+}}{\tilde{r}} \right)
\left( 1 -
  \frac{\tilde{r}_{-}}{\tilde{r}} \right)^{(1-\alpha^{2})/(1+\alpha^{2})}
   \label{eqn:funcdef}
\end{eqnarray}
with
\begin{equation}
 \tilde{r}_{\pm} = \frac{(1+\alpha^2) (M \pm \sqrt{
  M^{2} - \left( 1-\alpha^{2} \right) Q^{2}})}{\left(1 \pm \alpha^{2} \right)},
\; \;    \;
 a = \frac{2 (1 + \alpha^2) J}{ (1 + \alpha^2) \tilde{r}_{+} +
(1-\alpha^{2}/3) \tilde{r}_{-}}   .
  \label{eqn:parameteraprx}
\end{equation}
 This solution is valid only when $ a $ is sufficiently small.  Although
$ f(\tilde{r}) $ seems to diverge at $ \alpha = 1/\sqrt{3} $, $ \alpha = 1 $
and $ \tilde{r}_{-} = 0 $, $ f(\tilde{r}) $
approaches  a finite limiting value exists at each point.  \\
\hspc
 The angular velocity $ \OmH $ of this solution diverges for $ \alpha \geq
1/\sqrt{3} $ in the maximally charged limit, but vanishes for $ \alpha <
1/\sqrt{3} $.   There seems to exist a critical value
of the coupling constant, beyond which we expect that the superradiant
phenomena would change drastically (see \S. 2). Since the maximally charged
limit is no longer valid in the present approximation,
$ \alpha =1/\sqrt{3} $ may not be this critical value.
However, from the analysis of slowly rotating black holes we may still  find
some qualitative difference in the superradiant phenomena for a  large coupling
constant  as we shall see now.
\\ \hspc
 In this approximate solution, we can write down the radial part of  the
Klein-Gordon equation as
\begin{equation}
   \left[ \frac{d^2}{{d {r}^*}^{2}} + \left( \omega -  m
\Omega (\tilde{r}) \right)^{2} - {V}^2 (\tilde{r})
 \right] \chi ({r}^*) = 0
 \label{eqn:KG-slow}
\end{equation}
where
\begin{eqnarray}
 \Omega (\tilde{r}) & \equiv &
   \frac{a  f (\tilde{r})}{ R^{2} (\tilde{r})},\\
V^2 (\tilde{r}) & \equiv &
\frac{\Delta (\tilde{r})}{R^{2} (\tilde{r})}
\left[
\frac{l(l+1)}{R^{2} (\tilde{r})}  +
  \frac{1}{R (\tilde{r})}
\frac{d}{d \tilde{r}}
\left(
{\Delta(\tilde{r}) \over {R^2(\tilde{r})}}
\frac{d  R (\tilde{r})}{d \tilde{r}}
  \right)
\right],
\end{eqnarray}
and
$ {r}^* $ is defined by
\begin{equation}
 d {r}^* = \frac{ {R^2(\tilde{r})}}{\Delta (\tilde{r})} d \tilde{r}   ,
\label{eqn:tortoiseaprx}
\end{equation}
\hspc
Fixing  the small angular momentum $ J = 0.01 M^{2} $,
we change the charge $Q$. We have analyzed for $ \alpha = 0.5, 0.9, 1.1, 1.5,$
and $  2.0$  in addition to the cases of $ \alpha = 0,$ and $  \sqrt{3} $,
which were already analyzed in \S.2 but are included in the present
calculations to see the accuracy of the present approximation.
The results of the numerical calculations are shown in Fig.3.  \\
\hspc
 This calculation shows that when we increase the coupling constant, the
superradiant emission rates are enhanced. In particular, for  sufficiently
large $ \alpha $, they diverge in the maximally charged limit, although they
seem to approach to a finite value for the small $ \alpha $ case.
There seems to be a critical value of the coupling constant, below which the
emission rates are similar to the Kerr-Newman case, while
beyond which they diverge in the maximally charged limit.  Although
the present solution we used is only valid for a slowly rotating limit, the
comparison of the results with the exact solutions of $\alpha=0, $ and $
\sqrt{3}$ suggests that we may extrapolate the present analysis to
the case of rather highly charged black hole for which one may expect the
present approximation is broken down.   From the figure, we find that there is
a big change  near
$\alpha \sim 1$ in the behavior of the emission rates, which is consistent with
the previous analysis in \S. 2.
We can speculate that the critical value is 1, which is reduced from a
superstring model, as in the case of black-hole temperature\cite{CDBsol}.
\\\section{Discussion}
\hspc
 Here we shall discuss an evolution of rotating dilatonic black holes  created
in nearly extreme states. If a black hole has small charge initially, the
effect of the dilaton coupling is small because the dilatonic black hole in the
limit of $ Q \rightarrow 0 $ always approaches  the Kerr solution.
On the other hand, if we consider a highly charged black hole, the black hole
with  larger coupling constant $ \alpha $ than the critical value
$\alpha \sim1 $ will lose their angular momentum much faster than the black
hole with small $ \alpha $.  For example, the Kaluza-Klein black hole loses a
half of its initial angular momentum by the factor $ 10^{-10} $ shorter than
the Kerr-Newman does. (see Fig. 2).\\
\hspc
As for the  dependence of coupling constant, we may conclude that there is a
critical value below and beyond which
the qualitative behavior in the superradiance changes drastically.
The critical value would be unity, which is reduced from a superstring model.
\\
\hspc
In this paper, we have considered only the case of conserved charge.
If a discharge process exists, black holes with any $ \alpha $ will approach
quickly to the Kerr solution, as is pointed out in the Kerr-Newman case
\cite{Page}. Shiraishi \cite{SuperradQ} also pointed out that a nonrotating
black hole with larger $ \alpha $ will lose its charge faster than that with
small $ \alpha $.
Therefore we expect that total emission of energy  of black hole with larger $
\alpha $ is larger initially. But eventually, the difference will get much
smaller after some evolution. \\
\hspc
The contribution from the Hawking thermal radiation may become more important
than the superradiant effect during the evolution, because the temperature of
black hole with large $ \alpha $ is greater than that with small $ \alpha$ and
it diverges in the maximally charged limit.  If it is the case, our conclusion
in this paper would be changed.
As Holzhey and Wilczek \cite{Wilczek} pointed out, however, the Hawking
radiation may not be so important if the coupling constant
is larger than the critical value, because the potential barrier get very large
in the maximally charged limit and may prevent particles from passing through
it from the horizon to infinity.   This is now under investigation by numerical
calculation.
%%%%%%%%%%%%%%%%%%%%%%%%%%%%%%%%%%%%%%%%%%%%%%%%%%%%%%%%%%%%%%%%

{\bf Acknowledgment}\\
We would like to
thank T. Tachizawa for useful
discussions.
 This work was supported partially by the
Grant-in-Aid for Scientific Research  Fund of the
Ministry of Education, Science and Culture  (No.
06302021 and No. 06640412), and by Waseda University
Grant for Special Research Projects.

%%%%%%%%%%%%%%%%%%%%%%%%%%%%%%%%%%%%%%%%%%%%%%%%%%%%%%%%%%%%%%%
\vspace{2cm}
\begin{flushleft}
{\bf Figure Captions}
\end{flushleft}
\baselineskip .65cm

\vskip 0.1cm
\noindent
\parbox[t]{2cm}{\bf Fig. 1:\\~\\~\\~}\ \ \parbox[t]{14cm}{
The ranges of the angular momentum $J/M^2$ and the charge $Q/M$
for event horizons to exist in (1) the Kerr-Newman and (2) the Kaluza-Klein
black holes.}\\[1em]
\noindent
\parbox[t]{2cm}{\bf Fig. 2:\\~\\~\\~}\ \ \parbox[t]{14cm}{Comparison of the
superradiant emission rates  of the Kaluza-Klein black holes with those of the
Kerr-Newman black holes,  fixing the angular momentum $ J = 0.01 M^{2} $:
(1) and (2) are the emission rates of energy and angular momentum for the
Kaluza-Klein black hole, respectively, while
(3) and (4) are the same for the Kerr-Newman black hole.}\\[1em]
\noindent
\parbox[t]{2cm}{\bf Fig. 3:\\~\\~}\ \ \parbox[t]{14cm}{Superradiant emission
rate of the energy (a) and the angular momentum (b)
for a slowly rotating black holes with
(1) $ \alpha = 0 $ (the result by the exact solution is denoted by the
circles),
(2) $ \alpha = 0.5 $,
(3) $ \alpha = 0.9 $,
(4) $ \alpha = 1.1 $,
(5) $ \alpha = 1.5 $,
(6) $ \alpha = \sqrt{3} $  (the result by the exact solution is denoted by the
triangles) , and
(7) $ \alpha = 2.0 $.  The charge is normalized by the maximal value
$ Q_{max} = \sqrt{1+ \alpha^{2}} M $.
}\\
\newpage
\begin{figure*}[h]
 \psbox[height=21cm,width=14cm]{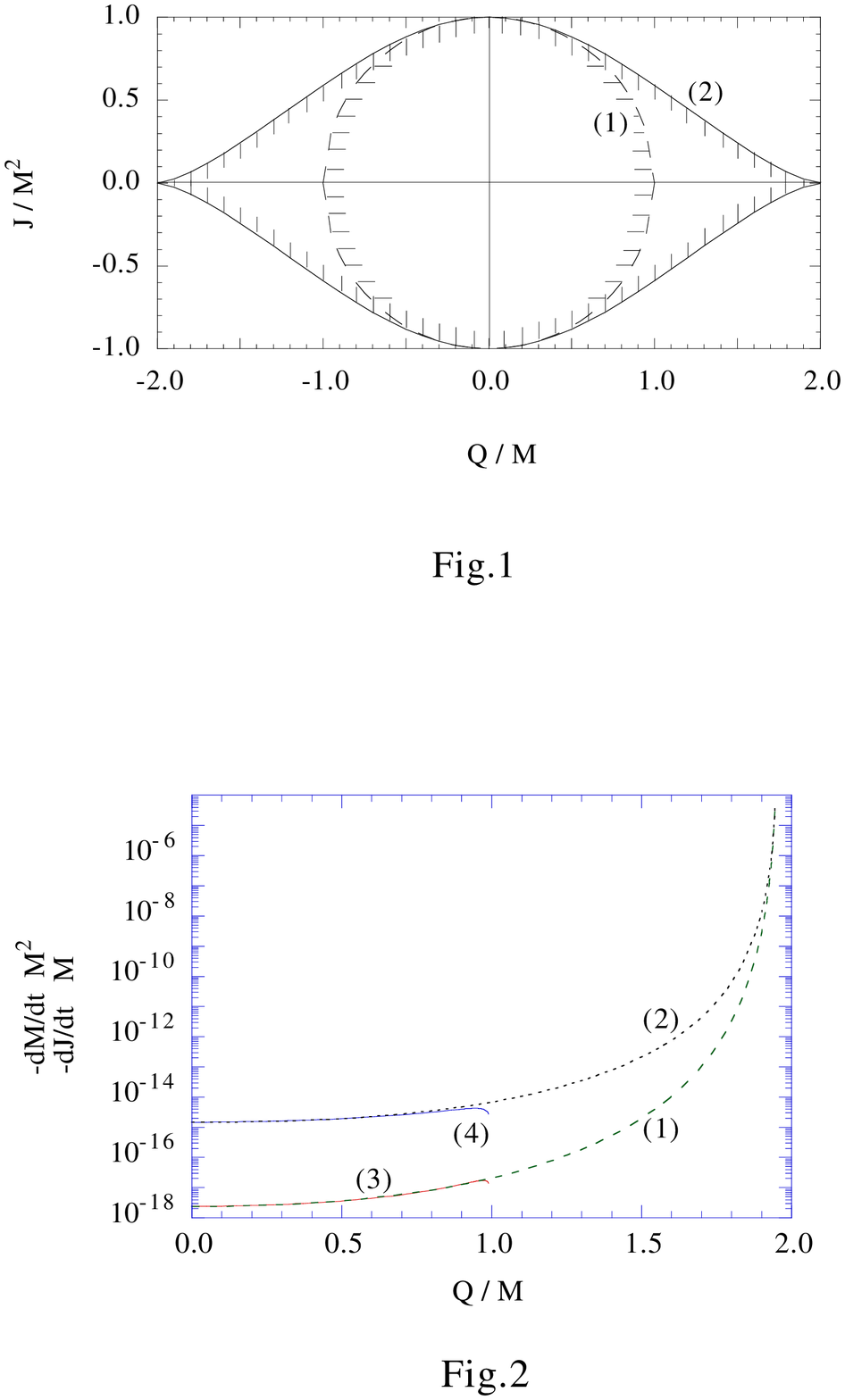}
\end{figure*}
\begin{figure*}[h]
 \psbox[height=21cm,width=14cm]{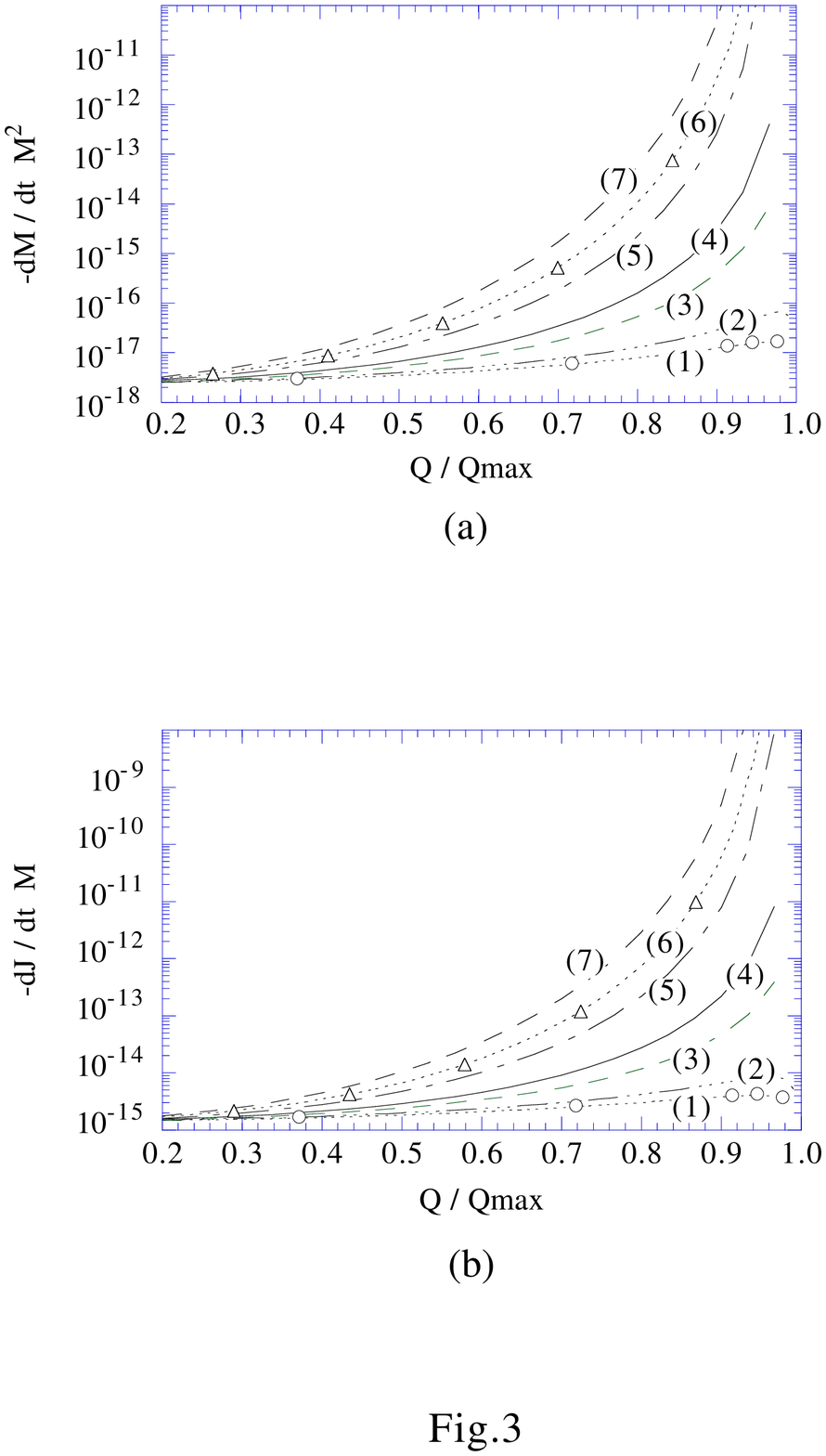}
\end{figure*}
\end{document}